\newcommand{\krakow}{Krak\'ow{}}
\documentclass{PoS}
\usepackage[]{graphicx}
\usepackage{array}
\usepackage{eurosym}
\title{The small size telescope projects for the Cherenkov Telescope Array}
\ShortTitle{The SST projects for CTA}
\author{\speaker{T. Montaruli}$^a$ for the CTA Consortium\footnote{Full consortium author list at http://cta-observatory.org} and
for the SST-1M sub-consortium:\\
\footnotesize{
W.~Bilnik$^{j}$,
J. B\l{}ocki$^{b}$,
L.~.Bogacz$^{d}$,
T~.Bulik$^{c}$,
F.~Cadoux$^{a}$,
A.~Christov$^{a}$,
M.~Cury{\l}o$^{b}$,
D.~della Volpe$^{a}$,
M.~Dyrda$^{b}$,
Y.~Favre$^{a}$,
A.~Frankowski$^{f}$,
\L{}. Grudniki$^{b}$,
M.~Grudzi{\'n}ska$^{c}$,
M.~Heller$^{a}$,
B.~Id{\'z}kowski$^{d}$,
M.~Jamrozy$^{d}$,
M.~Janiak$^{f}$,
J.~Kasperek$^{j}$,
K.~Lalik$^{j}$,
E.~Lyard$^{a}$,
E.~Mach$^{b}$,
D.~Mandat$^{k}$,
A.~Marsza{\l}ek$^{b,d}$,
J.~Micha{\l}owski$^{b}$,
R.~Moderski$^{f}$,
A.~Neronov$^{a}$,
J.~Niemiec$^{b}$,
M.~Ostrowski$^{d}$,
P.~Pa{\'s}ko$^{e}$,
M.~Pech$^{k}$,
A.~Porcelli$^{a}$,
E.~Prandini$^{a}$,
P.~Rajda$^{j}$,
M.~Rameez$^{a}$,
E.jr~Schioppa$^{a}$,
P.~Schovanek$^{k}$,
K.~Seweryn$^{e}$, 
K.~Skowron$^{b}$,
V.~Sliusar$^{g}$,
M.~Sowi{\'n}ski$^{b}$,
{\L}.~Stawarz$^{d}$,
M.~Stodulska$^{d}$,
M.~Stodulski$^{b}$,
S.~Toscano$^{a}$, 
I.~Troyano Pujadas$^{a}$, 
R.~Walter$^{a}$,
M.~Wi{\c e}cek$^{j}$,
A.~Zagda\'{n}ski$^{d}$,
K.~Zi{\c e}tara$^{d}$,
P.~{\.Z}ychowski$^{b}$.\\
a.
Universit\'e de Gen\`eve, Switzerland.
b.IFJ  PAN,
31-342 Krak{\'o}w, Poland.
c. Astronomical Observatory, University of Warsaw, 
 00-478 
Poland.
d. 
Jagiellonian University, 
30-244
\krakow, Poland.
e. CBK PAN,  
00-716 Warsaw, Poland.
f. Nicolaus Copernicus Astronomical Center, 
Warsaw, Poland.\\
g. Astronomical Observatory, 
University of Kyiv, 
Ukraine.
j. AGH University of Science and Technology, 
\krakow, Poland.
k. Institute of Physics, Czech Academy of Sciences, 
Prague, Czech Republic.
}}
\author{{T. Greenshaw$^{li}$ and H. Sol$^{po}$ for the GCT sub-consortium:}\\
\footnotesize{
A.~Abchiche$^{po}$,
J.-P.~Amans$^{po}$,
T.~Armstrong$^{du}$,
A.~Balzer$^{au}$,
D.~Berge$^{au}$,
J.J. Bousquet$^{po}$,
A.~Brown$^{du}$,
M.~Bryan$^{au}$,
G. Buchholtz$^{po}$,
P.~Chadwick$^{du}$,
H.~Costantini$^{mu}$,
G.~Cotter$^{ou}$,
M.~Daniel$^{li}$,
F. De Frondat$^{po}$,
J.L.~Dournaux$^{po}$,
D.~Dumas$^{po}$,
J.P. Ernenwein$^{mu}$,
G. Fasola$^{po}$,
A.~De Franco$^{ox}$,
J. Gaudemard$^{po}$,
J.~Hinton$^{mp}$,
J.-M.~Huet$^{po}$,
J.~Lapington$^{le}$,
P. Laporte$^{po}$,
S.J.~Nolan$^{du}$,
J.~Osborne$^{le}$,
S.~Rosen$^{le}$,
D.~Ross$^{le}$,
H.~Sol$^{po}$,
G.~Rowell$^{ad}$,
J.~Schmoll$^{du}$,
R.~Stuik$^{au}$,
P.~Sutcliffe$^{li}$,
J.~Sykes$^{le}$,
H.~Tajima$^{na}$,
R.~White$^{mp}$,
A. Zech$^{po}$.\\
ad. High Energy Astrophysics Group, University of Adelaide,
Australia.
na. Solar-Terrestrial Environment Laboratory, Nagoya University, 
464-8601 Japan.
du. 
Centre for Advanced Instrumentation, University of Durham,
DH1 3LE, UK.
mu. CPPM, 
CNRS/IN2P3, Marseille, France. 
ou. Oxford Astrophysics, 
OX1 3RH, UK.
au. GRAPPA, 
 University of Amsterdam, 
 1098 XH Amsterdam, Netherlands.
po. Observatoire de Paris, CNRS, 
F-92190, Meudon, France.
li. Dept. of Physics, 
Liverpool University, L69 7ZE, UK.
le. Dept. of Physics and Astronomy, University of Leicester, 
LE1 7RH,  UK.\\
mp. MPI f\"{u}r Kernphysik, 
69029 Heidelberg, Germany.
}}
\author{{G. Pareschi$^{1}$ for the ASTRI sub-consortium:}\\
\footnotesize{
E. Antolini$^{1}$,
L. A. Antonelli$^{1}$, 
D. Bastieri$^{1}$, 
G. Bellassai$^{1}$, 
C. Bigongiari$^{1}$,  
B. Biondo$^{1}$, 
M. Boettcher$^{2}$, 
A. Burtovoi$^{1}$, 
G. Bonanno$^{1}$,   
G. Bonnoli$^{1}$, 
P. Bruno$^{1}$,   
A. Bulgarelli$^{1}$, 
R. Canestrari$^{1}$,
M. Capalbi$^{1}$,
P. Caraveo$^{1}$, 
A. Carosi$^{1}$, 
E. Cascone$^{1}$, 
O. Catalano$^{1}$,  
P. Conconi$^{1}$, 
V. Conforti$^{1}$,  
G. Crimi$^{1}$,  
G. Cusumano$^{1}$, 
V. De Caprio$^{1}$,   
E.M. de Gouveia Dal Pino$^{3}$, 
M. Del Santo$^{1}$,
A. Di Paola$^{1}$, 
F. Di Pierro$^{1}$, 
A. Di Stefano$^{1}$,
I. Donnarumma$^{1}$,
D. Fantinel$^{1}$, 
C.E. Fermino$^{1}$,
V. Fioretti$^{1}$, 
M. Fiorini$^{1}$, 
D. Fugazza$^{1}$, 
S. Gallozzi$^{1}$,
D. Gardiol$^{1}$, 
C. Gargano$^{1}$, 
S. Garozzo$^{1}$, 
P. Gianmaria$^{1}$, 
F. Gianotti$^{1}$, 
S. Giarrusso$^{1}$, 
R. Gimenez$^{3}$, 
E. Giro$^{1}$, 
A. Giuliani$^{1}$, 
A. Grillo$^{1}$,  
D. Impiombato$^{1}$, 
S. Incorvaia$^{1}$,  
N. La Palombara$^{1}$, 
V. La Parola$^{1}$, 
G. La Rosa$^{1}$, 
L. Lessio$^{1}$, 
G. Leto$^{1}$, 
S. Lombardi$^{1}$,  
F. Lucarelli$^{1}$, 
M. C. Maccarone$^{1}$, 
A. Mangano$^{1}$, 
D. Marano$^{1}$, 
E. Martinetti$^{1}$, 
C. Melioli$^{3}$,
D. Messi$^{3}$,
M. Miraglia$^{1}$, 
T. Mineo$^{1}$,
G. Molrlino$^{1}$,
M. Munari$^{1}$, 
R. Nemmen$^{3}$, 
G. Occhipinti$^{1}$,
L. Perri$^{1}$, 
M. Perri$^{1}$,
G. Piano$^{1}$,
G. Rodeghiero$^{1}$, 
P. Romano$^{1}$,   
G. Romeo$^{1}$, 
A. Rubini$^{1}$, 
Fed. Russo$^{1}$, 
Fr. Russo$^{1}$,
P. Sangiorgi$^{1}$, 
B. Sacco$^{1}$,  
S. Sabatini$^{1}$,  
S. Scuderi$^{1}$,
J. Schwarz$^{1}$,  
A. Segreto$^{1}$,   
G. Sironi$^{1}$, 
G. Sottile$^{1}$,  
A. Stamerra$^{1}$, 
L. Stringhetti$^{1}$, 
C. Tanci$^{1}$,  
K. Tayabaly$^{1}$,
M. Tavani$^{1}$,
F. Tavecchio$^{1}$,
V. Testa$^{1}$, 
M. C. Timpanaro$^{1}$, 
G. Toso$^{1}$, 
G. Tosti$^{1}$, 
M. Trifoglio$^{1}$, 
G. Umana$^{1}$, 
S. Vercellone$^{1}$, 
R. Zanmar$^{1}$, 
L. Zampieri$^{1}$,
V. Zitelli$^{1}$
A. Zoli$^{1}$\\
$^1$ INAF \& the ASTRI collaboration.
$^2$ North-West University \& the ASTRI collaboration.
$^3$ Universidade de Sao Paulo \& the ASTRI collaboration.
}}
\abstract{The small size telescopes (SSTs), spread over an area of several square km, dominate the CTA sensitivity in the photon energy range from a few TeV to over 100 TeV, enabling for the detailed exploration of the very high energy gamma-ray sky. The proposed telescopes are innovative designs providing a wide field of view. Two of them, the ASTRI (Astrophysics con Specchi a Tecnologia Replicante Italiana) and the GCT (Gamma-ray Cherenkov Telescope) telescopes, are based on dual mirror Schwarzschild-Couder optics, with primary mirror diameters of 4 m. The third, SST-1M, is a Davies-Cotton design with a 4 m diameter mirror. Progress with the construction and testing of prototypes of these telescopes is presented. The SST cameras use silicon photomultipliers, with preamplifier and readout/trigger electronics designed to optimize the performance of these sensors for (atmospheric) Cherenkov light. The status of the camera developments is discussed. The SST sub-array will consist of about 70 telescopes at the CTA southern site. Current plans for the implementation of the array are presented.}

\FullConference{The 34th International Cosmic Ray Conference,\\
		30 July- 6 August, 2015\\
		The Hague, The Netherlands}

\begin{document}

\section{The CTA small size telescope projects}
\begin{figure}[hbt]
\centering
  \includegraphics[width=0.48\textwidth]{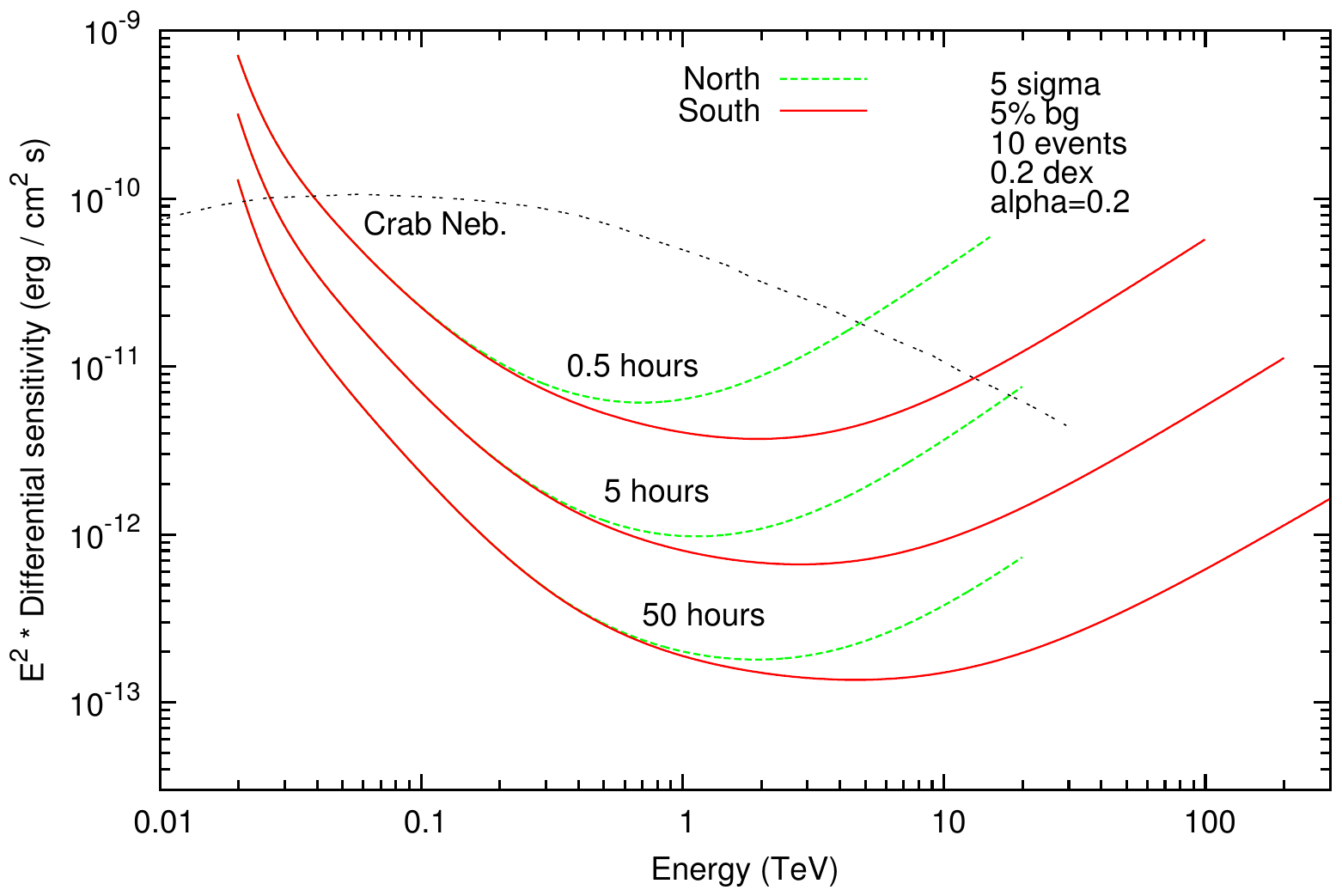}
   \includegraphics[width=0.48\textwidth]{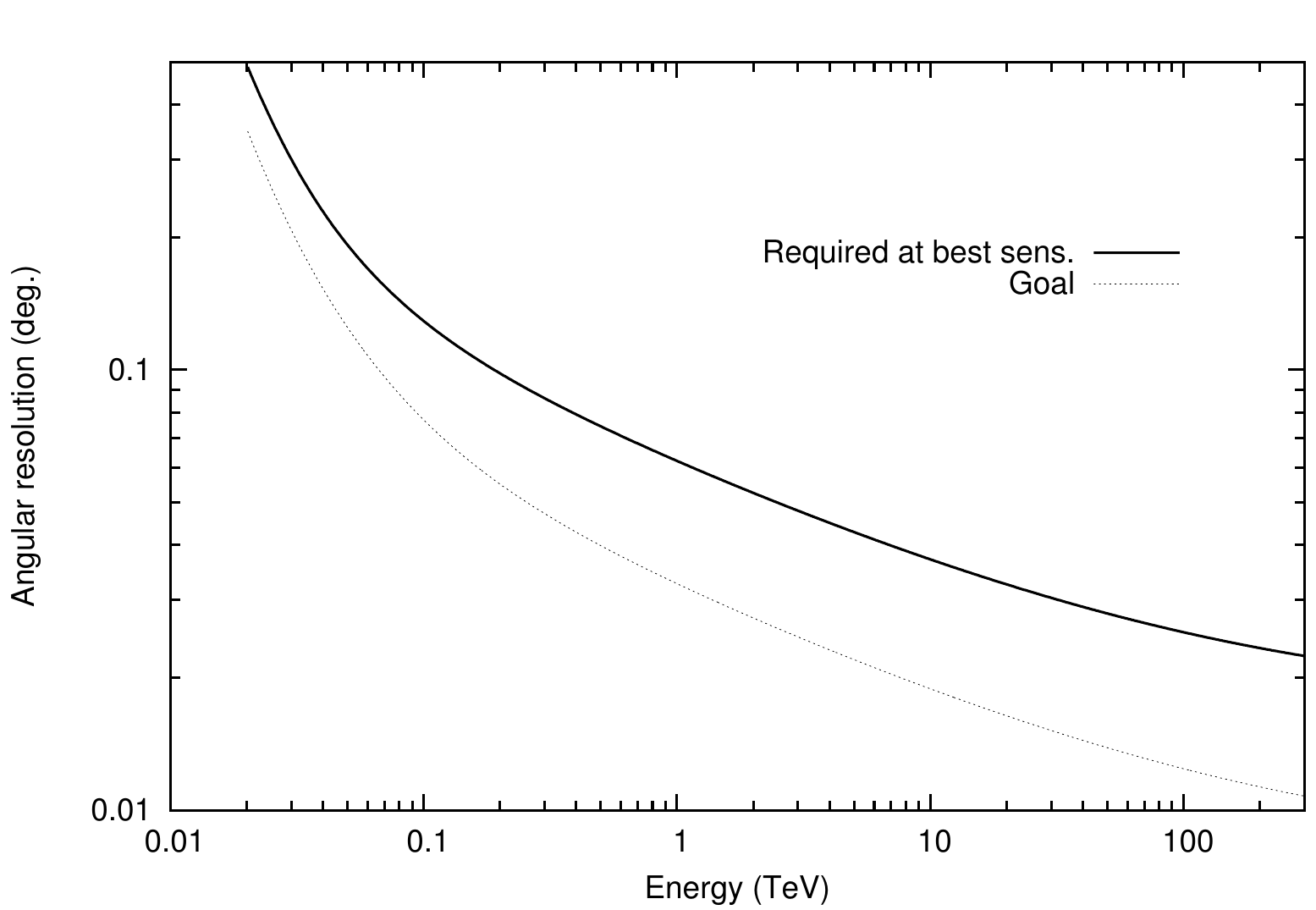}
  \caption{Left: Required dark-sky differential sensitivity ($5\sigma$ C.L.) for the CTA arrays, calculated in 5 bins per
energy decade, for different observation times and scaled with $E^2$.
 For comparison, the dashed black line indicates the Crab Nebula flux. Right:
Required and goal dark-sky angular resolution for the CTA southern array as a function of energy, showing the 68\%
containment radius of the point spread function.}
  \label{fig:requirements_sens}  
\end{figure}

The Cherenkov Telescope Array (CTA) will be the world's first open-access ground-based very high-energy gamma-ray observatory.
It will be composed of a northern and a southern array. The latter, planned to be located at the ESO (Armazones 2K) site in Chile or at Aar in Namibia,
will include many tens of small size telescopes (SSTs), spread over more than a square kilometer.
This large area sub-array is essential if the
observatory is to reach its required sensitivity in the photon energy range from about
5 to 300~TeV (see Fig.~\ref{fig:requirements_sens}-left). 

The goal light-collection area of the SST sub-array above 100~TeV is over 7~km$^2$.
Monte Carlo studies indicate that this can be achieved with an array of about 70 telescopes
at separations of about 250~m with 4~m diameter mirrors. Given the large amount of Cherenkov light generated by
multi-TeV gamma-rays, efficient triggering is possible even with mirrors of this relatively small size. Moreover,  the light pool is large enough to enable several telescopes to see each high-energy photon-induced shower with these telescope separations.
The required and goal angular resolutions of CTA are shown in Fig.~\ref{fig:requirements_sens}-right.
At 10 TeV, the angular resolution must be at least $0.04^\circ$ and $0.03^\circ$ at 100TeV. At these energies,  
CTA sensitivity is dominated by the SST section of the array.

The camera pixel charge resolution must be better than 30\% (15\%) above a signal of 
$\sim 20 (100)$ photoelectrons (p.e.). The charge resolution depends on the performance of the camera sensors and electronics
which must also provide a linear response over a charge range from 0 up to 2000~p.e.

Another important requirement for the SSTs is the field of view (FoV), which needs to be $> 8^{\circ}$.
This feature makes these
telescopes the first instruments capable of observing over such a large portion of the sky in the gamma-ray band, 
ensuring they are excellent tools for surveys and extended source observations.


Responding to the design challenges outlined above
has required considerable R\&D, which has resulted in instruments with innovative optics and cameras.
The use of silicon
photomultipliers (SiPMs), rather than conventional photomultipliers, allows the extension
of observations into periods with appreciable moonlight, as SiPMs do not suffer damage at high light
levels. This increases the SST sensitivity where observations are limited by the number of
signal gamma-rays observed, rather than background levels. The First G-APD Cherenkov Telescope
(FACT) \cite{FACT} has demonstrated that successful operation is possible even with full moon, as long as trigger
thresholds are adjusted accordingly. Further advantages of SiPMs are that they operate at modest bias
voltages, are lightweight and mechanically robust and exhibit only very slow aging. SiPM performance
has improved significantly in the last few years
and this has been accompanied by falling prices.

Two complementary solutions are proposed for the SSTs to contain costs while achieving the required
sensitivity: a single-mirror Davies-Cotton telescope with a SiPM-based camera (SST-1M) and two telescope
designs with dual-mirror Schwarzschild-Couder optics (SST-2M) and compact cameras, ASTRI
and GCT. The ASTRI camera is based on SiPMs, while the GCT is considering the use of SiPMs or multi-anode PMTs (MAPMs).
At the time the GCT camera prototypes were proposed, MAPMs were the cheaper option and the MAPM sensitivity in the $300 \div 350$\,nm wavelength range was higher than that of SiPMs. Improvements in the price and performance of SiPMs have now reversed this situation.

\begin{figure}[hbt]
\centering
  \includegraphics[width=0.9\textwidth]{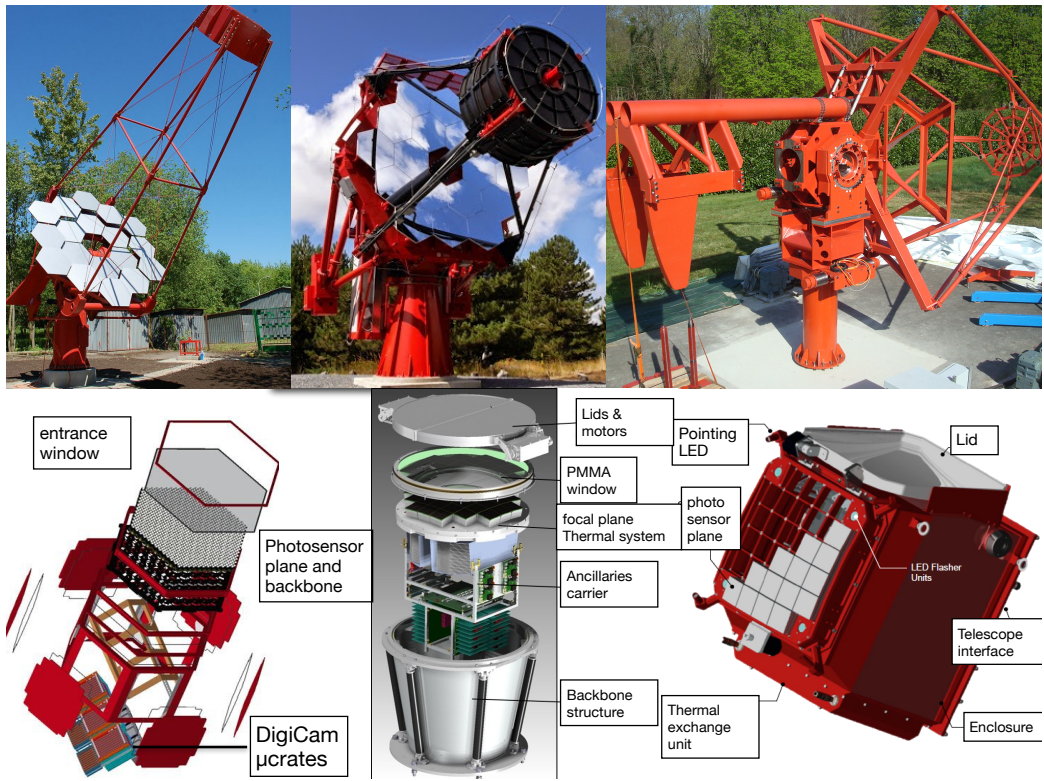}
  \caption{From left to right: The SST-1M structure inaugurated in Nov.~2013 in Krakow. Middle: The ASTRI prototype at Serra La Nave, inaugurated in Sep.~2014. Right: the  GCT structure in Paris in Apr.~2015. Prototype camera designs are shown below the telescopes.}
  \label{fig:telescopes}  
\end{figure}

Tab.~\ref{tab1} presents the main characteristics of the telescope designs and the pictures of the three prototypes and their
camera designs are shown in Fig.~\ref{fig:telescopes}.

The SST-1M, proposed by Swiss and Polish institutions, has a focal length of 5.6~m and a pixel size of $0.24^\circ$. The pixels are made of hexagonal
SiPMs (flat-to-flat dimension 10~mm) coupled to industrially produced light-collectors coated to enhance
reflectivity at small incidence angles for blue and near-UV light~\cite{light guides}. The camera has 1296 channels and is
about 0.9~m in diameter. It is sealed with a Borofloat window with a filter that rejects visible and IR light
with a wavelength above 540~nm, to reduce the effect of the night-sky background (NSB).

The dual mirror telescopes, ASTRI and GCT, have focal lengths of 2.15 and 2.28~m, respectively. The
increased complexity of the Schwarzschild-Couder optics with respect to that of the Davies-Cotton
design is counterbalanced by the reduced dimensions of the cameras. These have a diameter
below about 0.4~m and an angular pixel size of $0.15^\circ$ to $0.20^\circ$ for pixels
of $6 \times 6$ mm$^2$ to $7 \times 7$ mm$^2$.
 Both ASTRI camera and Compact High Energy Camera (CHEC) of GCR employ commercially
available arrays of sensors (monolithic arrays of SiPMs and MAPMs), providing about 2000 channels, and use 
cost-effective readout ASICs which have low
power consumption.

\begin{table}
\begin{tabular}{|l|c|c|c|}\hline
{\bf SST parameters }& {\bf ASTRI} & {\bf GCT} & {\bf SST-1M}\\ \hline
Eff. mirror collecting area (m$^2$)& 6 &6 & 6.47\\ 
Focal length (m) &2.15 & 2.28 &5.6\\
Field of view & $9.6^\circ$ & $9.2^\circ$ &$9.1^\circ$\\
On axis PSF (80\% $\gamma$ inclusion)& $0.17^\circ$ & $0.1^\circ$ & $0.08^\circ$ \\
Pixel size (mm) & 6.1 (square) & 7 (square) &  6 (hexagon)\\
Dish diameter (m)& 4 &4 &4\\
n. camera channels & 2368 & 2048 & 1296 \\
Camera diameter (m) & 0.40 &0.35& 0.88\\
Camera mass (kg) &55 & 45 & 186\\
Max. power camera + cooling (kW) &  0.8 & 0.9 & 2.9\\
Typical power consumption in 24 hrs (kWh) &41 & 30.1 & 57.8 \\
Data rate at 600 Hz (Gb/s)&  0.08&  2 & 0.2 for (80 ns readout window)\\
Readout window length (ns) & $12.5 \div 100$ & 96 &  $20 \div 2000$\\
Telescope mass (tons) &  20 &  7.8 &  8.6\\
 \hline
\end{tabular}
\caption{\label{tab1} The characteristics of the three proposed SST telescopes.
}
\end{table}

Multiple SST designs increase the prototyping and the CTA implementation
effort but nonetheless prototyping of new technologies such as the optics and the cameras of SSTs is an essential learning step. This situation developed in the preparatory phase as a result of funding
imperatives. Currently, the three projects aim to provide an in-kind contribution of at least 20 telescopes each to CTA, with as many common systems as possible to minimize
the impact on the effort needed for operation.
Once the projects pass the review process and in-kind contributions are funded, these telescopes will form the 70 SST sub-array at the southern site.

Coordination of the SST-1M, ASTRI and GCT projects is ensured by the SST Steering Committee. The major goal of this coordination is 
to increase the commonality of SST components and subsystems. For
example, the committee has ensured that the GCT and ASTRI cameras will adopt the same SiPMs for
their final design, and that the interface between the telescope and the cameras is the same, allowing
both cameras to be used on either dual mirror telescope. The committee is working towards further
unification including common infrastructure (e.g. foundations) and common telescope control hardware and software.
This strategy will keep the cost of the infrastructure and operation of the SST sub-array as low as possible.

\paragraph{SST-1M: The single mirror SST}
The SST-1M is particularly attractive since it has a simple, compact, and lightweight mechanical
structure with 8.6 tons total weight, straightforward to construct and produce, install, and maintain. 
The design, whose characteristics are in Tab.~\ref{tab1}, is stiff and solid, suitable for sites well above 2~km in altitude
and able to resist to the Chile site earthquake conditions. 
Synergy with the CTA Medium-Size Telescopes (MST) in terms of the
drive system components and the control software was achieved.

The prototype SST-1M telescope structure was installed at INP PAS in Krak\'ow in Nov.~2013 (see Fig.~\ref{fig:telescopes}, top left). Fourteen of the 18 glass mirror facets, of hexagonal shape, 78~cm side-to-side and with SiO$_2$ coating,  were installed in May 2015 and the camera
will be installed in fall 2015. The automatic mirror alignment system includes eighteen actuators. However, the prototype will allow evaluation of whether 
manual alignment (e.g. employed by FACT \cite{FACT}) is an option complying with the CTA maintenance requirement of less than 3.5~hrs per week, currently fully
satisfied by the SST-1M since cost would decrease. 

The prototype facets have been produced by different companies to compare their reflectivity, spot size and degradation in time. The 
baseline producer, the Armenian Galaktika company which produced the H.E.S.S. mirrors, offers the lowest
cost and produces the best spot size, but mirrors will require re-coatings during the CTA lifetime. 
Such facets produce a PSF, estimated by simulations that include information on the real prototype facets, of $0.07^\circ$ on-axis and $0.22^\circ$ at $4^\circ$ off-axis \cite{Rafal}.
DOTI, the producer of VERITAS facets, with substrate coating by Olomouc, member of SST-1M, provides another interesting solution with higher initial reflectivity above the
specification of 85\%. INAF glass cold-shaping mirrors are very attractive since they are lightweight but their spot size is at the limit of specifications (spot size with 80\% of photons $D_{80}$ = 0.8 mrad at 2f distance). More details on the structure and the optical system are presented in~\cite{SST-1Mproc,karol}. 

The camera \cite{SST-1Mcamera} (Fig.~\ref{fig:telescopes} bottom left) is lightweight and has 1296 channels. Each pixel is made of a hexagonal SiPM of 95 mm$^2$ area coupled with an open industrially produced light guide with
high reflectivity for wavelengths lower than 400~nm and almost normal incidence. Wavelengths above 540~nm are filtered out by the 3.3~mm-thick Borofloat window. The large silicon area of sensors required quite an effort on the
frontend electronics to achieve required linearity and charge resolution. The slow control board, part of the photosensor plane, ensures a self-calibrating camera with a feedback system which tunes the
bias voltage according to temperature changes.
The DigiCam (based on the FlashCam approach) full digitizing readout system of the camera guarantees deadtime-less data acquisition (DAQ), high throughput and
high flexibility in trigger selections~\cite{rajda}.

Simulations indicate that the SST-1M design satisfies the required performance for SSTs \cite{Rafal}. 
The camera calibration strategies and calibrations with muons are described in \cite{proc_calib,proc_calib1}.

\paragraph{ASTRI: a dual mirror SST}
A prototype (Fig.~\ref{fig:telescopes} top center) has been developed and installed at the Serra La Nave Observatory on Mt. Etna (Sicily) and is being tested and qualified. 
This first prototype telescope design has some margins in terms of performance and costs. The prototype construction provides a verification of the system concept. It helps understanding where improvements can be made in an updated design that will be used for the realization of an array of SST precursors (9 units) that is proposed to be installed at the CTA southern site from mid-2016~\cite{bib:Vercellone}.

The ASTRI telescope prototype \cite{bib:Canestrari} has been designed to match the physical pixel size provided by SiPM sensors ($\sim 6$~mm) to the required angular pixel size of the SST. A primary mirror (M1) of 4.3~m diameter is used, with a focal ratio of about 0.5. The mirror profiles are aspherical, with substantial deviation (a few tens of mm in sag) from the main spherical component. The optical system has a plate scale of 37.5 mm/$^\circ$ and an angular pixel size of $\sim 0.17^\circ$. With the adopted design, a light collection efficiency higher than 80\% within the dimension of the pixels over the entire FoV of $10^\circ$ and an average effective area of about 5~m$^2$ are achieved.

The ASTRI telescope adopts an alt-az mount in which the azimuth axis allows rotation over $\pm$270$^\circ$. The mirror dish is mounted on the azimuth fork which allows a rotation around the elevation axis from -5$^\circ$ to +95$^\circ$. The mast structure that supports the secondary mirror and the camera is fixed on the mirror dish. In order to balance the torque due to the overhang of the optical tube assembly with respect to the horizontal rotation axis, counterweights are also supported by the mirror dish.
The primary mirror is based on a set of 18 hexagonal panels with 850~mm face-to-face dimension \cite{bib:Canestrari_mirr}. These are produced via cold slumping of glass foils. The secondary aspherical mirror is a glass monolithic shell of $\sim 2$~m diameter, produced via hot slumping.

The azimuth motion is obtained by means of a couple of ratio-gear-motors, with their pinions engaged with a bearing rack. A brake is installed on one of the ratio-gear motors. The M1 dish structure (made of a ribbed steel plate of about 40~cm thickness) is attached to the azimuth fork using two preloaded tapered roller bearings, one on each of the fork's arms. The elevation bearings are installed on the top of the azimuth fork and permit the rotation around the M1 elevation axis. The supports for the mirror segments are attached to the dish, each of which includes a single and a double axis actuator and a bearing. The structure for supporting the monolithic secondary mirror is located at the upper end of the mast. The orientation of the telescope is determined using encoders located on the azimuth and altitude axes. Finite element analysis (FEA) has been used to evaluate the performance of the system. The lowest frequency eigenmode of the structure is 4.5~Hz. FEA has also been carried out to determine the effects on the telescope of temperature gradients, wind loads and earthquakes \cite{bib:Canestrari}.

The ASTRI prototype camera \cite{bib:Catalano} (see Fig.~\ref{fig:telescopes} bottom center ), adopts SiPM as photosensors. It is under construction and will be mounted on the ASTRI prototype telescope in fall 2015, with the aim of performing astronomical Cherenkov gamma-ray observations for calibration purposes.
The sensor chosen for the ASTRI prototype is the Hamamatsu S11828-334 SiPM, consisting of $4 \times 4$ pixels of roughly $3 \times 3$~mm$^2$ each. Four Hamamatsu pixels are grouped together to form a pixel with a physical size of 6.1$\times$6.1~mm$^2$, matching the required angular size (future ASTRI cameras will use larger pixels that are now becoming available). Four of the Hamamatsu devices are put together to form a unit. Four such units then form a module called a Photon Detection Module (PDM). This module is composed of 16 Hamamatsu devices and has dimensions of $56\times 56$~mm$^2$. The PDMs are constructed by plugging the Hamamatsu devices into connectors attached to a printed circuit board (PCB). Under each PCB there is a small temperature sensor, allowing monitoring of the SiPM temperature, providing a route to stabilize the gain.

The Front End Electronics (FEE) boards of the camera supply the power to the SiPM, perform the readout and form the first trigger signals. The CITIROC (Cherenkov Imaging Telescope Integrated Read Out Chip) ASIC is adopted. 
The Back-End Electronics (BEE) is based on an FPGA and local memory to provide interfaces to the CTA DAQ, 
and to the CTA trigger and clock distribution system. There is also circuitry to provide the various DC voltages needed to power the elements of the camera.

The total height of the camera is about 30~cm. A lid system is used to protect the sensors. As the focal plane of the 2M telescopes is convex, with a radius of curvature of 1~m, the PDMs are attached to a precisely machined curved plate. Under the sensor plane is positioned the support structure to which further electronics boards and the cooling system are attached.
\paragraph{The Gamma-ray Cherenkov Telescope and its camera} 
The GCT telescope structure has been designed to 
allow easy assembly on the remote CTA southern site. It consists of a 
tower
supporting the telescope's drive
system, the bearings, gears and motors which allow motion in the
azimuth and altitude directions.
The same drive systems 
are used for both axes.
The alt-az system holds the optical assembly and the counterweight support. The
former consists of the primary dish
and the masts that hold the secondary mirror and camera,
while the latter holds the counterweights, which can be moved during 
assembly to ensure the telescope is balanced.
The camera is held beneath the secondary mirror 
on a swiveling mount that provides easy access 
for installation and maintenance.

The mirrors are constructed using either polished and coated aluminum, or glass.
The former will be used on the prototype.
The primary mirror is formed of 6 trapezoidal segments, each of which can be mounted
from ground level by rotating the mirror support structure
about the telescope's axis. Following installation, locking of this support ensures 
the required
stability. The secondary mirror is constructed of 6 petals,
mounted on the telescope as a monolithic unit. 
All primary segments and the secondary mirror are mounted via actuator systems
which allow alignment of the mirrors and focussing of the telescope. 
Mirrors will be mounted on the prototype GCT on the Meudon site of the Paris 
Observatory, shown in Fig.~\ref{fig:telescopes}, top right, in summer 2015.

FEA has
demonstrated that the required mirror location precision 
is achieved for the full range of CTA operating
conditions and that the telescope can
survive both the highest wind speeds
and the seismic conditions of the CTA southern site (Aar or Paranal). 
Studies of the prototype structure at Meudon will be used to confirm these results.

The GCT prototype MAPM-based Compact High-Energy Camera
(CHEC-M) is shown in Figure~\ref{fig:telescopes} (bottom right).
Behind the MAPMs is a pre-amplifier which amplifies and shapes the
signals from the sensors.
Different pre-amplifiers are used for CHEC-S, the camera prototype with SiPMs. 
They ensure appropriate SiPM signal pre-processing before
entering the readout chain. The readout is based on the ASIC
TARGET 5(7) for CHEC-M (CHEC-S)~\cite{TARGET}.
The TARGET chip samples and digitizes the
incoming waveforms at a rate of 1\,Gs/s.
Each TARGET chip has 16 parallel input channels and is placed on
a board which provides the power necessary for the chip and the
associated sensors. The board also steers the readout and some control
functions via an FPGA. 
Four such boards
are grouped together to form a TARGET module, which provides readout
for an $8 \times 8$ array of pixels, attached via the pre-amplifiers
to its front end. The attachment system allows the
compensation of the 1\,m radius of curvature of the focal plane on
which the sensors lie, so that the TARGET modules can be placed in a
rectilinear crate inside the camera body. The rear end of
each TARGET module is attached to the backplane, a
multi-layer printed circuit board which receives the trigger signals from the
TARGET modules and combines them in an FPGA to form the camera
trigger. Following a trigger, formed by requiring signals above
threshold in a number of neighboring trigger pixels (each of which is
a $2 \times 2$ block of camera pixels), readout is initiated and
steered by two DAQ boards which pass the waveform
provided by TARGET to the remote CTA central DAQ in the control room.

A CHEC-M prototype has been assembled in Leicester. This is 
complete except for the
backplane, which will be added in summer 2015. Tests of the prototype,
using a dummy backplane which routes the data from the TARGET 
modules directly to the DAQ boards, have demonstrated the functionality 
of all aspects of the camera, with the exception of the camera trigger. 

Next steps for GCT include the installation and testing 
of CHEC-M on the prototype telescope, with its mirrors, in 
autumn 2015.
CHEC-S will be assembled and tested in the laboratory on a 
similar timescale, before installation and test on the telescope in early 
2016. Any needed design changes following these tests will
be made and then three pre-production telescopes and cameras 
manufactured, shipped to the CTA southern site, assembled 
and commissioned. Following acceptance tests, a
further 32 telescopes and 35 cameras 
(providing 3 spares) will be built and installed.

\acknowledgments
We gratefully acknowledge support from the agencies and organizations 
listed under Funding Agencies at this website: http://www.cta-observatory.org/.

\end{document}